\documentclass[preprint,review,12pt]{elsarticle}

\usepackage{hyperref} 

\usepackage{setspace}
\usepackage{amsmath,amsfonts,amssymb,amsthm,mathrsfs,bm}
\usepackage{graphicx} 
\usepackage[usenames,dvipsnames]{xcolor}
\usepackage{xspace}

\usepackage{soul}

\usepackage{booktabs}
\usepackage{multirow}
\usepackage{makecell}

\usepackage[version=4]{mhchem}
\usepackage{siunitx}

\usepackage{tikz}
\usetikzlibrary{
    3d,
    calc,
    math,
    shadows,
    patterns,
    external,
    arrows.meta,
    backgrounds,
    shapes.geometric,}
\usepackage{tikz-3dplot} 
\usepackage{pgfplots}
\usepackage{pgfmath}
\usepgfplotslibrary{groupplots}

\pgfplotsset{compat=1.18}
\definecolor{warm1}{HTML}{FFD099}
\definecolor{warm2}{HTML}{FF9209}
\definecolor{warm3}{HTML}{FF6C22}
\definecolor{cool1}{HTML}{3081D0}
\definecolor{cool2}{HTML}{2E8B57}

\usepackage{siunitx}

\newcommand{\paren}[1]{\left(#1\right)}
\newcommand{\set}[1]{\left\{#1\right\}}
\newcommand{\bracket}[1]{\left[#1\right]}

\newcommand{\modelname}{\texttt{BCAE-VS}\xspace}

\newcommand{\bcae}{\texttt{BCAE}\xspace}
\newcommand{\bcaetwod}{\texttt{BCAE-2D}\xspace}
\newcommand{\bcaeht}{\texttt{BCAE-HT}\xspace}
\newcommand{\bcaeplus}{\texttt{BCAE++}\xspace}

\newcommand{\dseg}{D_\text{seg}}
\newcommand{\dreg}{D_\text{reg}}

\newcommand{\mink}{\texttt{MinkowskiEngine}\xspace}

\newcommand{\ada}{\texttt{NVIDIA RTX\textsuperscript{TM} 6000 ADA}\xspace}

\newcommand{\chip}{\ada}
\newcommand{\driver}{\texttt{driver 555.42.02}\xspace}
\newcommand{\torch}{\texttt{PyTorch 2.4.0}\xspace}
\newcommand{\cuda}{\texttt{CUDA 12.2}\xspace}
\newcommand{\Title}{Variable Rate Neural Compression for Sparse Detector Data}

\makeatletter
\def\ps@pprintTitle{%
  \let\@oddhead\@empty
  \let\@evenhead\@empty
  \let\@oddfoot\@empty
  \let\@evenfoot\@oddfoot
}
\makeatother
\newif\ifeditingfigure
\editingfigurefalse

\begin{document}

\begin{frontmatter}



\title{\Title}


\author[label1]{Yi Huang}
\author[label1]{Yeonju Go}
\author[label1]{Jin Huang}
\author[label3]{Shuhang Li}
\author[label1]{Xihaier Luo}
\author[label5]{Thomas Marshall}
\author[label1]{Joseph D.~Osborn}
\author[label1]{Christopher Pinkenburg}
\author[label1]{Yihui Ren}
\author[label1,label4]{Evgeny Shulga}
\author[label1]{Shinjae Yoo}
\author[label1,label2]{Byung-Jun Yoon}

\affiliation[label1]{
    organization={Brookhaven National Laboratory},
    addressline={PO Box 5000}, 
    city={Upton},
    state={NY},
    postcode={11973}, 
    country={USA}
}
\affiliation[label2]{
    organization={Texas A\&M University},
    addressline={3128 TAMU}, 
    city={College Station},state={TX},
    postcode={77843}, 
    country={USA}
}
\affiliation[label3]{
    organization={Columbia University},
    addressline={538 West 120th Street}, 
    city={New York},
    state={NY},
    postcode={10027}, 
    country={USA}
}
\affiliation[label4]{
    organization={Stony Brook University},
    addressline={100 Nichols Road}, 
    city={Stony Brook},
    state={NY},
    postcode={11794}, 
    country={USA}
}
\affiliation[label5]{
    organization={University of California, Los Angeles},
    addressline={475 Portola Plaza}, 
    city={Los Angeles}, 
    state={CA},
    postcode={90095},
    country={USA}
}

\begin{abstract}
High-energy large-scale particle colliders generate data at extraordinary rates, reaching up to one terabyte per second in nuclear physics and several petabytes per second in high-energy physics. Developing real-time high-throughput data compression algorithms to reduce data volume and meet the bandwidth requirement for storage has become increasingly critical. Deep learning is a promising technology that can address this challenging topic. At the newly constructed sPHENIX experiment at the Relativistic Heavy Ion Collider, a Time Projection Chamber (TPC) serves as the main tracking detector, which records three-dimensional particle trajectories in a volume of a gas-filled cylinder. In terms of occupancy, the resulting data flow can be very sparse reaching $10^{-3}$ for proton-proton collisions. 
Such sparsity presents a challenge to conventional learning-free lossy compression algorithms, such as SZ, ZFP, and MGARD. 
In contrast, emerging deep learning-based models, particularly those utilizing convolutional neural networks for compression, have outperformed these conventional methods in terms of compression ratios and reconstruction accuracy. However, research on the efficacy of these deep learning models in handling sparse datasets, like those produced in particle colliders, remains limited. Furthermore, most deep learning models do not adapt their processing speeds to data sparsity, which affects efficiency.
To address this issue, we propose a novel approach for TPC data compression via \emph{key-point identification} facilitated by \emph{sparse convolution}. Our proposed algorithm, \modelname, achieves a $75\%$ improvement in reconstruction accuracy with a $10\%$ increase in compression ratio over the previous state-of-the-art model. Additionally, \modelname manages to achieve these results with a model size over two orders of magnitude smaller. Lastly, we have experimentally verified that as sparsity increases, so does the model's throughput. 
Our code along with the pretrained models and dataset used for model development are available at \url{https://github.com/BNL-DAQ-LDRD/NeuralCompression_v3} and \url{https://zenodo.org/records/10028587}, respectively. 
\end{abstract}




\end{frontmatter}


\allowdisplaybreaks
\section{Introduction}
\label{sec:introduction}

\begin{figure}[ht]
    \centering
    \ifeditingfigure
        \tikzsetnextfilename{detector_assembly}
        \def\fontsz{15}
        \def\scale{.65}
        \input{fig/sphenix_assembly}
    \else
        \includegraphics{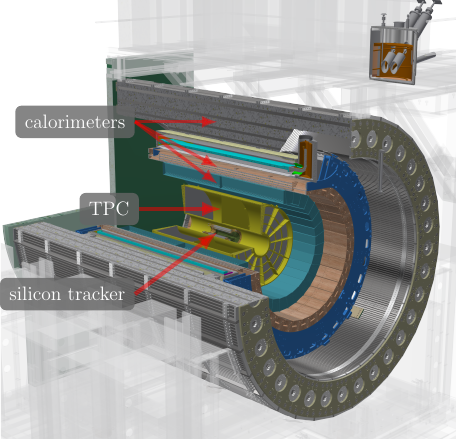}
    \fi
    \caption{\textbf{sPHENIX detector assembly. Time projection chamber (TPC) contributes the majority of data.}}
    \label{fig:sphenix}
\end{figure}

\subsection{Background}

High-energy particle colliders, such as the Large Hadron Collider (LHC)~\cite{evans_lhc_2008} and the Relativistic Heavy Ion Collider (RHIC)~\cite{sPHENIX_TDR}, are critical tools for probing the fundamental building blocks of matter. 
In these experiments, protons and heavy ions are accelerated to velocities approaching the speed of light and providing collisions at extremely high energies (for RHIC nucleon-nucleon center of mass energy is $\SI{200}{\giga\eV}$ in heavy ion collisions up to and $\SI{510}{\giga\eV}$ in proton-proton collisions). These complex interactions generate a multitude of subatomic particles (thousands of particles in the central heavy ion collisions), enabling scientists to explore the deepest questions about the nature of the universe. To capture the products of collisions, the interaction point is surrounded by the state of art particle detectors. 
Recently constructed sPHENIX experiment~\cite{sPHENIX_TDR} at RHIC focuses on studying the microscopic properties of strongly interacting matter, such as the quark-gluon plasma, a state of matter that existed in the early universe.

To achieve this, sPHENIX employs an array of advanced particle detectors (\autoref{fig:sphenix}). The Time Projection Chamber (TPC)~\cite{Hilke:1302071} is the primary tracking detector. The TPC acts as a three-dimensional (3D) camcorder, continuously recording the trajectories of charged particles as they pass through its detection volume. As particles move through the TPC, they ionize the gas molecules, creating clusters of charge. Clusters drift in the electric field to the readout pads and induce electrical signals. These signals (hits) are registered by $160$~thousand detector channels. Both the position and time of a hit are captured by one of the $40$~million voxels in a 3D grid of the active detection volume. The system then digitizes these signals, transforming them into analog-to-digital conversion (ADC) values. The TPC readout digitizes $3.2$~trillion voxels per second and generates terabits per second of data after zero suppression.

Traditionally, collider experiments have relied on a level-1 trigger system to manage the massive data output by selecting only the most significant collision events for storage, while discarding less valuable data. However, next-generation collider experiments, including sPHENIX, are moving towards streaming data acquisition (DAQ) systems~\cite{sPHENIX_BUP23,Bernauer:2022moy,AbdulKhalek:2021gbh}, which aim to continuously record a large fraction to \emph{all} collision events. While this approach maximizes the physics output, it presents a critical challenge: handling the immense data volume in real time.

To reduce the TPC data volume while preserving key information, efficient data compression techniques are essential. This paper introduces a deep neural network-based lossy compression algorithm designed to meet the high-throughput demands of streaming DAQ systems. Moreover, unlike specialized algorithms tailored specifically for TPC data~\cite{rohr2020gpu}, our approach is \emph{data-driven} and makes no assumptions about the underlying physics, allowing it to generalize to other datasets with similar characteristics, such as those for the future electron-ion collider~\cite{AbdulKhalek:2021gbh,Bernauer:2022moy}.

\subsection{Related Work and Research Gaps}

Conventional scientific lossy compression methods are motivated in fields such as climate science, fluid dynamics, cosmology, and molecular dynamics. 
These methods have been driven by the necessity to handle large datasets produced by distributed and high-fidelity simulations in a high-performance computing environment. 
For example, the error-bounded SZ compression algorithm~\cite{di_fast_2016,tao_significantly_2017,liu_optimizing_2022} has proven effective for compressing data from climate science and cosmology simulations. Other examples include the ZFP method~\cite{lindstrom_fixed-rate_2014}, which was developed specifically for hydrodynamics simulations, and the MGARD method~\cite{ainsworth_multilevel_2019,liang_mgard_2022}, which was designed to compress turbulent channel flow and climate simulation data.

While there is extensive research on general-purpose lossy compression, few existing methods are optimized for highly sparse data. Although the aforementioned compression algorithms have demonstrated reasonable performance with TPC data, none of them are specifically tailored for this type of input. In this context, a specialized neural network-based compression model, the \textit{Bicephalous Convolutional Auto-Encoder} (\bcae), introduced in~\cite{huangTPCCompression} and later refined in~\cite{huang2023fast}, has been shown to outperform traditional methods in terms of both compression ratio and reconstruction accuracy for sparse TPC data.

However, despite the \bcae's superior performance, its design treats the sparsity of TPC data as a challenge to overcome rather than as an opportunity to improve compression efficiency. The \bcae model compresses data by mapping the input array to a \emph{fixed-size} code, regardless of the input’s occupancy (i.e., the fraction of non-zero elements). This results in inefficiencies: when the code size is large enough to accommodate inputs with high signal occupancy, there is wasted space for sparser inputs; conversely, when the code size is optimized for sparser inputs, denser inputs suffer from information loss during compression.

\subsection{Challenges and Solutions}
\noindent\textbf{Challenge 1.~Occupancy-based variable compression ratio.} 
While TPC data on average exhibits low occupancy (around $10\%$), the actual occupancy can vary significantly. For instance, in the dataset used for model development, occupancy ranges from less than $5\%$ to over $25\%$. Therefore, there is a need for compression algorithms capable of producing larger codes for denser inputs and smaller codes for sparser ones.


\noindent\textbf{Solution.} 
We propose \modelname, a bicephalous autoencoder that provides a variable compression ratio based on TPC data occupancy. The design of \modelname is motivated by the hypothesis that a trajectory can be reliably reconstructed from a subset of signals. Consequently, \emph{data compression can be achieved by selectively down-sampling signals rather than resizing the input array.}
To implement this, \modelname's encoder predicts the importance of each signal in the input array for trajectory reconstruction. Signals with higher importance are treated as analogous to the \emph{key points} that anchor shapes in computer vision tasks such as action and gesture recognition~\cite{avola20223d,lin2020human,you2020keypointnet,moskvyak2021keypoint}. Compression is achieved by saving only the coordinates and neural representations of these key points.

\noindent\textbf{Challenge 2.~Leveraging sparsity for throughput.} Despite the impressive reconstruction accuracy of \modelname, implementing its encoder with conventional (dense) convolution operations will lead to constant computation time disregarding input data with changing sparsity and complexity.

\noindent\textbf{Solution.} 
We use \emph{sparse convolution} for implementing \modelname's encoder. With a manageable runtime overhead, the sparse convolution-powered encoder of \modelname generates output only for the signals with input exclusively from the signals and skips all matrix multiplications with all-zero operands. As the overhead and the number of non-zero values vary with occupancy, \modelname achieves a variable throughput that increases significantly as occupancy decreases. 

\subsection*{\emph{\textbf{Main Contributions:}}}
\begin{itemize}
    \item \textbf{Introduction of \modelname:} We propose \modelname, an improved \bcae model enabling Variable compression ratio for Sparse data. This model aims to enhance both compression efficiency and reconstruction accuracy by selectively down-sampling signals rather than reducing the size of the input array.
    \item \textbf{Improved reconstruction performance:} \modelname achieves a $75\%$ improvement in reconstruction accuracy and a $10\%$ higher compression ratio on average compared to the most accurate \bcae model. 
    \item \textbf{Use of sparse convolution for high throughput:} To address the computational inefficiencies of conventional convolution with high sparsity data, \modelname employs sparse convolution. This approach processes only the relevant signals, significantly reducing the computational overhead associated with matrix multiplications involving all-zero operands. 
\end{itemize}
\section{TPC Data}
\label{sec:data}

\subsection{Overview}
\label{subsec:dataet}

\begin{figure}[ht]
    \centering
    \ifeditingfigure
        \tikzsetnextfilename{tpc_diagram}
        \resizebox{.6\textwidth}{!}{\input{fig/tpc_diagram}}
    \else
        \resizebox{.6\textwidth}{!}{\includegraphics{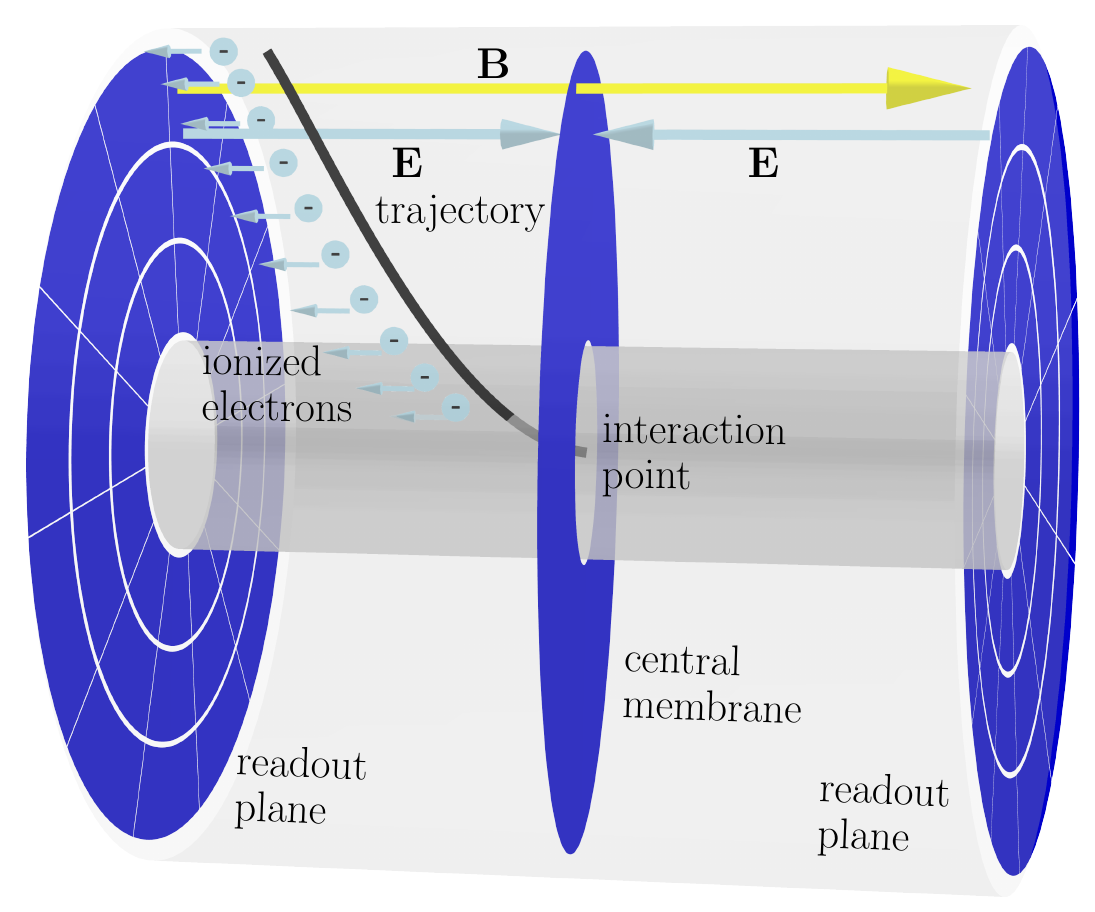}}
    \fi
    \caption{\textbf{A schematic view of sPHENIX TPC.}}
    \label{fig:tpc_diagram}
\end{figure}

\autoref{fig:tpc_diagram} shows a schematic view of sPHENIX TPC. sPHENIX TPC is a cylindrical gas-filled drift chamber. Working gas mixture is $75\%$ argon (\ce{Ar}), $20\%$ carbon tetrafluoride (\ce{CF4}), and $5\%$ isobutane (\ce{C4H10}). Collisions happen in the center of the TPC. Charged particles produced in the collisions traverse the volume of the TPC at a speed close to the speed of light. As charged particles travel, they ionize the gas atoms along trajectories and create trails of electrons and ions. Extremely uniform electric field $E=\SI{400}{\volt\per\centi\meter}$ provides uniform drift velocity for electron clouds drift towards the readout planes. The positions and arrival times of the electrons are recorded by the sensors arranged in concentric layers on the readout planes. Gas ionization by a particle depends only on its velocity and the square of the charge, thus the density of the electron cloud allows to distinguish different particles, e.g.~protons, kaons, and pions. The chamber is placed within the magnetic field of a superconducting magnet, which provides a magnetic field $B=\SI{1.4}{T}$ collinear with the electric field. Trajectory of the particle is bent by the magnetic field along the axial direction. 
This bending allows for the determination of the momenta and charges of the particles based on the curvature of their trajectory.

\begin{figure}[ht]
    \centering
    \ifeditingfigure
        \tikzsetnextfilename{tpc_full_wedge_track}
        \resizebox{\textwidth}{!}{\input{fig/tpc_full_wedge}}
    \else
        \resizebox{\textwidth}{!}{\includegraphics{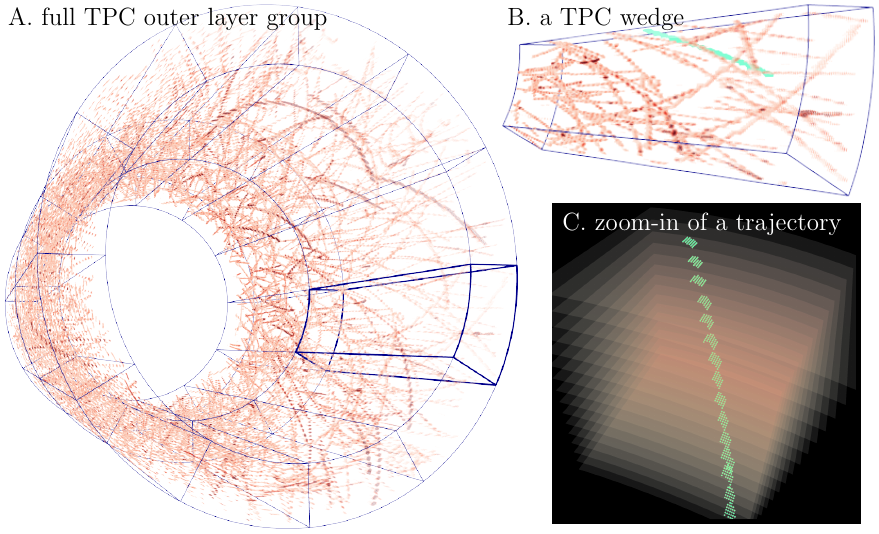}}
    \fi 
    \caption{\textbf{Visualization of the outer layer group TPC data of an Au$+$Au collision event with  $0$-$10$\% centrality and \SI{170}{\kilo\hertz} pileup.}}
    \label{fig:dataset}
\end{figure}

The sPHENIX TPC readout plane has three groups of layers: inner, middle, and outer. Each layer group produces a readout of a 3D (axial, radial, and azimuthal) array of ADC values. Each array has $16$ layers along the radial direction and $249$ temporal sampling points (for one TPC drift time window) along the axial direction on each side of the TPC, representing time evolution. The number of readout pads along the azimuthal direction increases from inner to outer layers. The number is constant within a layer group and it is $1152$, $1536$, and $2304$, respectively. In this study, we will focus on the $16$ layers of the outer group of the TPC with $2304$ readout pads along the azimuthal direction. Each element in the (layer, pad, temporal) array is mapped to a \emph{voxel} in the 3D space. We demonstrate a full azimuthal outer layer group readout in \autoref{fig:dataset}A (experiment $3060$, event $8$ in the test split of the dataset). 

To match the subdivision of the readout electronics in the TPC readout chain, the data from a layer group has the same division. It has $24$ equal-size non-overlapping sections: $12$ along the azimuthal direction ($30$ degrees per section) and $2$ along the axial direction (separated by the transverse plane passing the collision point). We call these sections \emph{TPC wedges}. \autoref{fig:dataset}B shows a zoom-in view of one of the wedges from the outer layer group. Each outer-layer wedge is an array of dimensions $16$, $192$, and $249$ in radial, azimuthal, and axial directions, respectively. All ADC data from the same wedge will be transmitted to the same group of front-end electronics, after which a real-time lossy compression algorithm could be deployed. Therefore, TPC wedges are used as the direct input to the deep neural network compression algorithms.

\subsection{Dataset and Preprocessing}

To train a compression algorithm, we used a dataset containing $1310$ simulated events for central $\sqrt{s_{\mathrm{NN}}}=200$~GeV Au$+$Au collisions with \SI{170}{\kilo\hertz} pileup collisions\footnote{Zenodo linked: \url{https://zenodo.org/records/10028587}}. The data were generated with the HIJING event generator~\cite{Wang:1991hta} and Geant4 Monte Carlo detector simulation package~\cite{Allison:2016lfl} integrated with the sPHENIX software framework~\cite{sPHENIX_Software}. 

The simulated TPC readout (ADC values) from these events are represented as $10$-bit unsigned integers $\in[0, 1023]$. To reduce unnecessary data transmission, a \emph{zero-suppression} algorithm has been applied. All ADC values below $64$ are set to be zero as large fraction of them represent noise. This zero-compression makes the TPC data sparse at about $10.8\%$ occupancy of non-zero values. A voxel that has a non-zero ADC value is a \emph{signal voxel}. A voxel with a $0$ ADC value is a \emph{non-signal voxel}. 

We divide the $1310$ total events into $1048$ events for training and $262$ for testing. Each event contains $24$ outer-layer wedges. Thus, the training partition contains $25152$ TPC outer-layer wedges, while the testing portion has $6288$. 

Finally, as trajectory coordinates must be interpolated from neighboring sensors using the ADC values, it is important to preserve the relative ADC ratio between the sensors. Hence, for this study, we work with \emph{log ADC values} (defined as $\log_2(\text{ADC} + 1)$) instead of the raw ADC values. A log ADC value is a float number $\in[0., 10.]$. As zero-suppression is implemented at $64$ for this dataset, and all nonzero log-ADC values exceed $6$. 
Zero-suppression makes the distribution of ADC bimodal and extremely skewed. 


\textbf{Remark:} This distribution presents a significant challenge for neural network-based algorithms since neural network tends to work well when input data is distributed approximately normally~\cite{alanazi2020simulation,hashemi2019lhc}. In \autoref{subsec:bcae} we will show one approach to handle the discontinuity in the data distribution with a compression autoencoder equipped with two collaborating decoders. We will discuss the limitation of this approach and provide a better solution in \autoref{subsec:bcae_spoi} with sparse convolution. As a result, the sparsity of the data is no longer an obstacle to overcome but rather an advantage that we can leverage.  
\section{Method}
\label{sec:method}

\subsection{Bicephalous Convolutional Autoencoder}
\label{subsec:bcae}

\begin{figure}[ht]
    \centering
    \ifeditingfigure
        \tikzsetnextfilename{bcae}
        \resizebox{.9\textwidth}{!}{\input{fig/bcae}}
    \else 
        \resizebox{.9\textwidth}{!}{\includegraphics{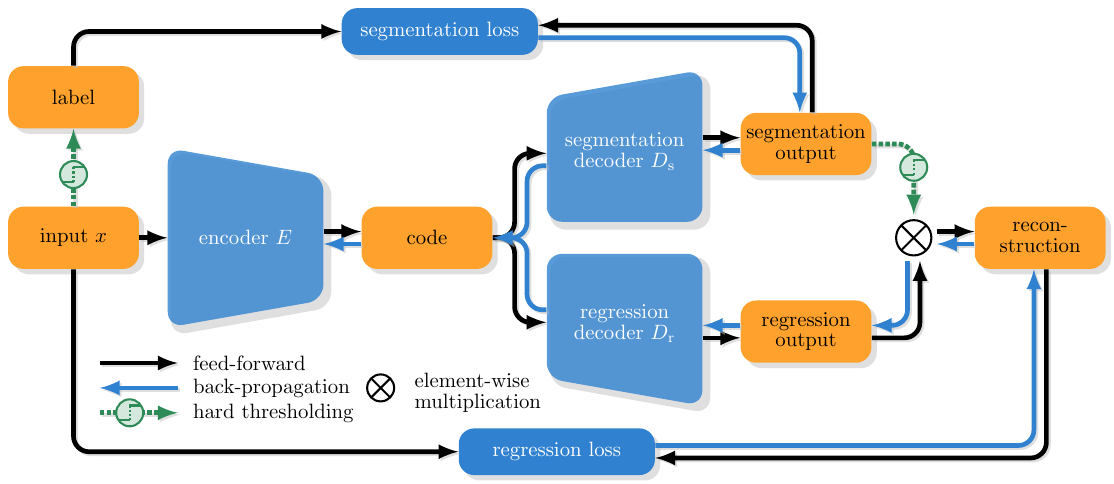}}
    \fi 
    \caption{\textbf{Bicephalous convolutional autoencoder}.}
    \label{fig:bcae}
\end{figure}

To address the problem caused by the difficult distribution of log ADC value, Bicephalous Convolutional Autoencoder~(\bcae) was proposed in~\cite{huangTPCCompression} and later optimized in~\cite{huang2023fast}. A \bcae is an autoencoder with two decoders -- one for segmentation and the other for regression (See \autoref{fig:bcae}). During training, a code significantly smaller than the input is generated by the encoder and fed to the decoders. The segmentation decoder, $\dseg$, will output a score $s\in(0, 1)$ for each voxel in the input. The weights of $\dseg$ are updated by comparing $s$ to the true label of the input -- $1$ for a signal voxel and $0$ for a non-signal voxel. The regression decoder, $\dreg$, will output a value $r$ for each voxel. Given a threshold $s_0 \in (0, 1)$, the predicted value ($\hat{r}$) at each voxel is set to $r$ if $s > s_0$ and to $0$ if otherwise. The weights of $\dreg$ are then updated by comparing $\hat{r}$ to the true input value of the voxel. The weights of the encoder $E$ are updated with the combined gradient information from both decoders. During inference, we only run the encoder part and save the code to storage for later use.  

The design of \bcae is motivated by two considerations: reconstruction accuracy and efficiency. First, the segmentation decoder is tasked to tackle the discontinuity in the input data distribution so that the regression decoder can focus solely on approximating the ADC values. Second, having two decoders does not affect the compression (inference) throughput as only the encoder will be executed during compression. From the first consideration, we can see that a better solution with respect to reconstruction accuracy might be to have two single-tasking encoder-decoder pairs. However, since the encoding will run in real-time in deployment, it is crucial to keep the encoder simple and lightweight as much as possible for the high TPC data rate. Additionally, as demonstrated in~\cite{huang2023fast}, the possible adverse effect of a simple encoder could be compensated to some extent by a more complicated decoder. 

Although \bcae has been shown to outperform existing non-neural network-based compression algorithms in reconstruction accuracy at equal or higher compression ratio~\cite{huangTPCCompression}, the approach does not leverage the sparsity of the input.
First, its dense convolution incurs static computation costs despite varying input sparsity. 
Second, to produce a significantly smaller code, the down-sizing in the encoder breaks the sparsity of the input. This makes it hard for the decoders to reconstruct a trajectory at its correct location, especially in higher occupancy areas (see \autoref{fig:reconstruction}).
Finally, for \bcae, the same code length is used regardless of data sparsity. 
This means that if a chosen code length is suitable for sparse and relatively simple data, it will become insufficient for denser and more complex ones.
\subsection{\modelname for Key-Point Identification}
\label{subsec:bcae_spoi}

\autoref{fig:dataset}C shows a close-up view of a typical trajectory formed by a streak of signal patches. The localized and seemingly redundant signal voxels of the same patch suggest the possibility of reconstructing the trajectory with only a fraction of them. As a starter, we performed a sanity check by randomly masking out $50\%$ of signal voxels and used a neural network to reconstruct the original input. This compression by random sampling does not even require a learnable encoder network! The reconstruction mean squared error (MSE) (in raw ADC) is $95$, which is much smaller than the $218$ reported in~\cite{huangTPCCompression}.

These results suggest that if randomly selected signal voxels can perform effectively in reconstruction, a carefully \emph{selected} subset of signal voxels should yield even better performance. Such a subset is analogous to the \emph{key points} in computer vision, to which location a geometric shape is anchored. This concept led to the development of \modelname whose encoder \emph{compresses by down-selecting signal voxels rather than down-sizing the whole input array}. 

To achieve compression by down-selecting, an encoder assigns an importance score to each signal voxel, retaining only those with higher importance and discarding the rest. However, implementing such an encoder with standard (dense) convolutions requires running 3D convolutions that maintain the input dimension, which can be relatively slow~\cite{huang2023fast}. This is where \emph{sparse convolution} becomes a potentially better alternative, as it can compute solely for the signal voxels and use input exclusively from them. 

This section is structured as follows: \autoref{subsubsec:sparse_bcae_model} explains the operation of \modelname. \autoref{subsubsec:sparse_conv} contrasts sparse convolution with dense convolution. In \autoref{subsubsec:random_threshold}, we explore using random thresholding for identifying key signal voxels. Finally, \autoref{subsubsec:losses} addresses the balance between compression and reconstruction in \modelname.
 
\subsubsection{The Working of \modelname}
\label{subsubsec:sparse_bcae_model}

\begin{figure*}[ht]
    \centering
    \ifeditingfigure
        \tikzsetnextfilename{me}
        \def\fontsz{10}
        \resizebox{\textwidth}{!}{\input{fig/me}}
    \else
        \resizebox{\textwidth}{!}{\includegraphics{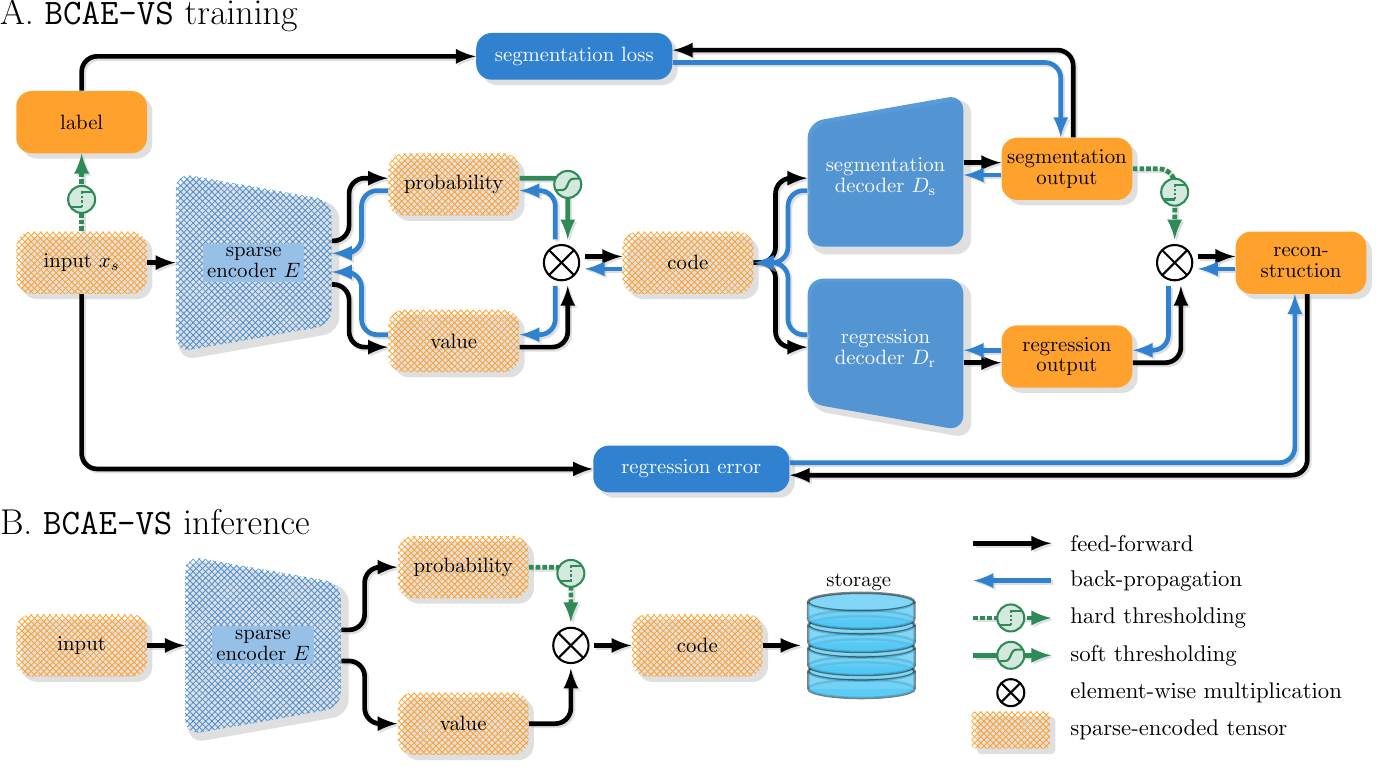}}
    \fi 
    \caption{\textbf{Bicephalous autoencoder with sparse encoder for key-points identification.}}
    \label{fig:me}
\end{figure*}

To describe \modelname in more detail, let us first fix some notations. We will use $x$ to denote a $n$-dimensional array and $x_s$ to denote the sparse form of $x$ with only the location and value of the non-zero entries specified. Although $x$ and $x_s$ represent the same input, the different notations emphasize the fact that zeros (non-signal voxels) do not occupy any space in $x_s$ while $x$ contains all the zeros. 

As demonstrated in \autoref{fig:me}, for each entry in a TPC wedge $x_s$, the encoder $E$ of \modelname produces two outputs, importance score $p$ and value $v$. The importance score output helps us determine the contribution of signal voxel in accurate reconstruction and the value is what we will save to the storage should the voxel be selected. There are two reasons for letting $E$ produce $v$ instead of using the original (raw or log) ADC value for reconstruction (as we did in the ``encoder-free'' compression). First, since we will have to save a value for a selected voxel anyway, an adjusted value produced by a trained neural network may be more helpful than the original (raw or log) ADC value. Second, the gain in inference speed from having one less output channel is marginal according to our experiment (at least for the specific sparse convolution implementation we chose for this work).

During the training of \modelname, we pass importance score $p$ through a soft mask function with an element-wise random threshold to get a soft mask $m$. The element-wise multiplication of $v$ and $m$ form the input to the decoders. For inference (compression only), the random soft thresholding will be replaced by a random hard thresholding. More details on the soft mask and random thresholding will be discussed in \autoref{subsec:bcae_spoi}.

\subsubsection{Sparse Convolution}
\label{subsubsec:sparse_conv}
\newcommand{\kh}{k_\text{h}}
\newcommand{\kw}{k_\text{w}}
\newcommand{\Kh}{K_\text{h}}
\newcommand{\Kw}{K_\text{w}}
For symbolic simplicity, we explain sparse convolution with input with two spatial dimensions, height and width. Generalization to a higher dimensional input is straightforward. We use $s$ and $d$ to denote the stride and dilation of a convolution, and $\Kh$ and $\Kw$ to denote the kernel window for the height and width direction, respectively. For example, for convolution with kernel size $3$, we have $\Kh=\Kw=[-1, 0, 1]$. We use $i$, $o$, $h$, $w$, $\kh$, and $\kw$ to iterate through the input and output channels, height and width, and the kernel, respectively. The output $Y$ can be calculated from the weight $W$, bias $\mathbf{b}$, and input $X$ as follows:
\begin{align}
    Y_{o, h, w} &= \sum_{i, \kh, \kw} W_{o, i, \kh, \kw}X_{i,\,s h + d\kh,\,s w + d\kw} + \mathbf{b}_o \tag*{}\\
    &= \sum_{\kh, \kw}\underbrace{\sum_{i}W_{o, i, \kh, \kw}X_{i,\,s h + d\kh,\,s w + d\kw}}_{\text{matrix multiplication}} + \mathbf{b}_o \label{eq:matmul}
\end{align}
For a dense convolution, the summation $\sum_{\kh, \kw}$ is taken over $\mathcal{K}=\Kh \times \Kw$ and hence we cannot avoid the matrix multiplication in \autoref{eq:matmul} even if the input vector $X_{i,\,sh + d\kh,\,sw + d\kw}$ is an all-zero vector. However, for a sparse convolution, the summation is taken over
\begin{align*}
    \mathcal{K}_{h, w} = \set{(\kh, \kw)\left|\paren{s{h} + d{\kh},\,s{w} + d{\kw}} \in \mathcal{S}\right.}\subset\mathcal{K}
\end{align*}
where $\mathcal{S}$ is the collection of coordinates in a sparse input. The map $(h, w)\mapsto\mathcal{K}_{h, w}$ is commonly referred to as the \emph{kernel map} in the sparse convolution literature. 

There are two types of sparse convolution: normal and submanifold sparse convolution. Let $\mathcal{N}_{h, w} = \set{\paren{s{h} + d{\kh},\,s{w} + d{\kw}}\left|(\kh, \kw) \in \mathcal{K} \right.}$ be the neighborhood of $(h, w)$ with respect to the kernel, normal sparse convolution computes for a coordinate $(h, w)$ if $\mathcal{N}_{h, w}\cap\mathcal{S}$ is not empty, while submanifold sparse convolution only computes for $(h, w) \in \mathcal{S}$. Since normal sparse convolution will introduce new entries to $\mathcal{S}$ every time the convolution with kernel size $>1$ is applied, and breaks the sparsity, we use \emph{submanifold sparse convolution} for the \modelname encoder.  

As we can see from the discussion above, while sparse convolution can reduce runtime by avoiding computations on zero input, it incurs overhead from computing kernel maps. Consequently, sparse convolution is more efficient at lower sparsity but loses its advantage as sparsity decreases. Additionally, the efficiency of sparse convolution varies with the specific implementation. For this work, we use the \mink library from NVIDIA\footnote{\url{https://github.com/NVIDIA/MinkowskiEngine}}, as it provides the highest throughput on the hardware used in this research. Details on throughput are available in \autoref{subsec:throughput}.

\subsubsection{Random Thresholding}
\label{subsubsec:random_threshold}

\begin{figure}[ht]
    \centering
    \ifeditingfigure
        \tikzsetnextfilename{soft_mask}
        \resizebox{\textwidth}{!}{\input{fig/soft_mask}}
    \else
        \resizebox{\textwidth}{!}{\includegraphics{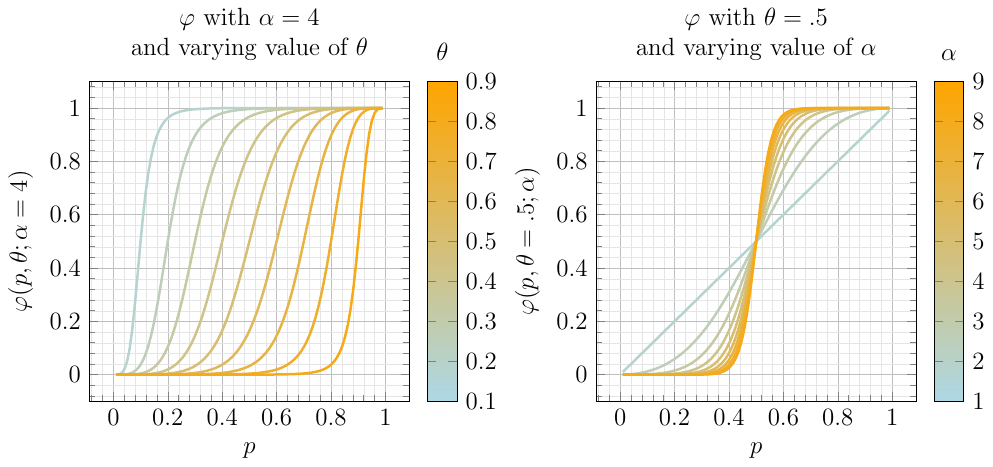}}
    \fi 
    \caption{\textbf{The soft mask function.}}
    \label{fig:soft_mask}
\end{figure}

\begin{figure}[ht]
    \centering
    \ifeditingfigure
        \tikzsetnextfilename{thresholding}
        \resizebox{\textwidth}{!}{\input{fig/thresholding}}
    \else 
        \resizebox{\textwidth}{!}{\includegraphics{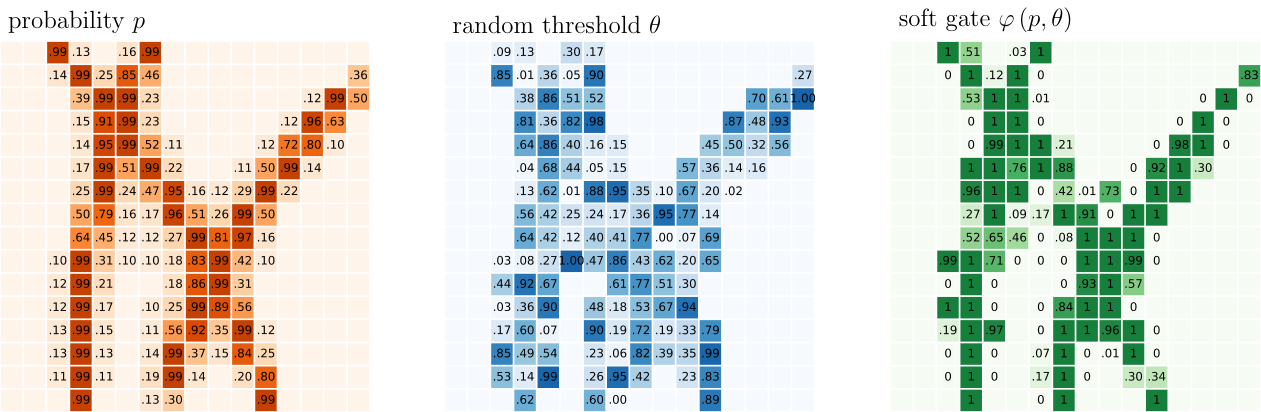}}
    \fi 
    \caption{\textbf{Random thresholding.} For each signal voxel, the encoder will calculate an importance score $p$ (left). A threshold $\theta$ is then generated following $\texttt{Uniform}(0, 1)$ for each voxel (middle). A soft mask is obtained by $\varphi(p, \theta)$ (right).}
    \label{fig:threshold}
\end{figure}

The problem of key-point identification in TPC data compression can be thought of as a ``self-supervised classification'' problem. It is a classification problem since we want to predict a label $1$ for a signal voxel crucial for accurate reconstruction and a $0$ for a non-essential one. It is a self-supervised problem since we do not have a ground truth label to match but have to infer the label by observing the effect of keeping or discarding a signal voxel on reconstruction. 

For this goal, we designed the random thresholding approach that works as follows. 
For a signal voxel with predicted importance score $p$, we generate a random number $\theta\sim\texttt{Uniform}(0, 1)$, and calculate a soft mask with the function defined in \autoref{eq:gate} 
\begin{equation}
    \label{eq:gate}
    \varphi(p, \theta;\alpha, \varepsilon) = \texttt{sigmoid}\paren{\alpha\bracket{\texttt{logit}(p, \varepsilon) - \texttt{logit}(\theta, \varepsilon)}}
\end{equation}
The $\varepsilon$ in the \texttt{logit} function is used to improve numerical stability ($p$ or $1-p$ less than $\varepsilon$ will be replaced with $\varepsilon$). We set $\varepsilon=10^{-8}$ for this study. Plots of $\varphi$ with varying threshold $\theta$ and $\alpha$ are shown in \autoref{fig:soft_mask}. We can see that $\varphi$ is a differentiable approximation to the step function with a step at $\theta$. For a fixed $\theta$, $\alpha$ increases the steepness of $\varphi$ when making the transition from $0$ to $1$. For this study, we set $\alpha=4$.


\subsubsection{Balancing Reconstruction with Compression}
\label{subsubsec:losses}
\newcommand{\lcomp}{\mathcal{L}_{\text{comp}}}
\newcommand{\lrecon}{\mathcal{L}_{\text{recon}}}
\newcommand{\lseg}{\mathcal{L}_{\text{seg}}}
\newcommand{\lreg}{\mathcal{L}_{\text{reg}}}
\newcommand{\cseg}{\lambda_{\text{seg}}}
\newcommand{\ccomp}{\lambda_{\text{comp}}}
\newcommand{\lowerprob}{l_{\text{prob}}}
To train \modelname, we need to balance the reconstruction accuracy with compression. We do this by adjusting the reconstruction loss $\lrecon$ and compression loss $\lcomp$.

The reconstruction loss $\lrecon$ is the weighted sum of the segmentation loss $\lseg$ and the regression loss $\lreg$. We use focal loss~\cite{lin2017focal} for segmentation since it is designed to deal with datasets with unbalanced classes (TPC data is sparse and hence has a high percentage of voxels in the class $0$). We use mean absolute error ($L_1$) for regression following the convention in~\cite{huang2023fast}. The reconstruction loss $\lrecon(x)$ is defined as $\cseg\lseg(x) + \lreg(x)$ with $\cseg$ adjusted by the end of each epoch with the same method as discussed in~\cite[section 2.5]{huang2023fast} 

Since \modelname compresses by identifying key points, we want the average of $p$, denoted as $\mu$, to be small, so that only the valuable signal voxels can get a high importance score. However, since there is no need for the average importance score to go arbitrarily small, we also set a lower bound $\lowerprob$ so that the average importance will be penalized more lightly when it is close to $\lowerprob$, and will no longer be penalized if it goes below $\lowerprob$. Hence, the compression loss function of \modelname is defined as
\begin{equation*}
    \lcomp\paren{x, \lowerprob} = \left\{
    \begin{array}{ll}
        {\mu(x_s)}\paren{\mu(x_s) - \lowerprob} & \mu(x_s) > \lowerprob\\
        0 & \mu(x_s) \leq \lowerprob
    \end{array}\right.
\end{equation*}

The overall loss $\mathcal{L}(x)$ of \modelname is defined as $\ccomp\lcomp\paren{x_s, \lowerprob} + \lrecon(x)$. In this study, we set $\lowerprob=.1$ and $\lambda_{\text{comp}}=30$. 
\section{Results}
\label{sec:results}

\subsection{Reconstruction Accuracy}
\label{subsec:reconstruction_accuracy}


\begin{table}[ht]
\centering
\caption{Comparing compression algorithm with dense and sparse convolutions.}
\label{tab:main_result}

\resizebox{\textwidth}{!}{
\begin{tabular}{lcccccccc}
    
    \toprule
    & & \multicolumn{5}{c}{reconstruction performance}  &  \multicolumn{2}{c}{efficiency} \\
    \cmidrule(l{.2em}r{.2em}){3-7}\cmidrule(l{.2em}r{.2em}){8-9}
    model & \makecell{comp.\\ratio $\uparrow$} & $L_1\downarrow$ & $L_2\downarrow$ & PSNR $\uparrow$ & recall $\uparrow$ & precision $\uparrow$ & \makecell{encoder\\size} & throughput $\uparrow$ \\
    
    \midrule
    \texttt{BCAE-2D} & $31$ & $.152$ & $.862$ & $20.6$ & $.907$ & $.906$ & $169$k & $\mathbf{9.6}$\bf{k} \\
    \texttt{BCAE-HT (3D)} & $31$ & $.138$ & $.781$ & $20.8$ & $.916$ & $.915$ & $9.8$k & $\mathbf{9.6}$k\\
    \texttt{BCAE++ (3D)}  & $31$ & $.112$ & $.617$ & $21.4$& $.936$ & $.934$ & $226$k & $3.2$k \\
    \texttt{\modelname} & $\mathbf{34}$ & $\mathbf{.028}$ & $\mathbf{.089}$ & $\mathbf{26.0}$ & $\mathbf{.988}$ & $\mathbf{.996}$ & $\mathbf{382}$ & $5.6$k \\
    \bottomrule
\end{tabular}
}

    
    

\end{table}

\begin{figure}[ht]
    \centering
    \ifeditingfigure
        \tikzsetnextfilename{metric_reg}
        \def\scale{.86}
        \input{fig/metric_reg}
    \else
        \includegraphics{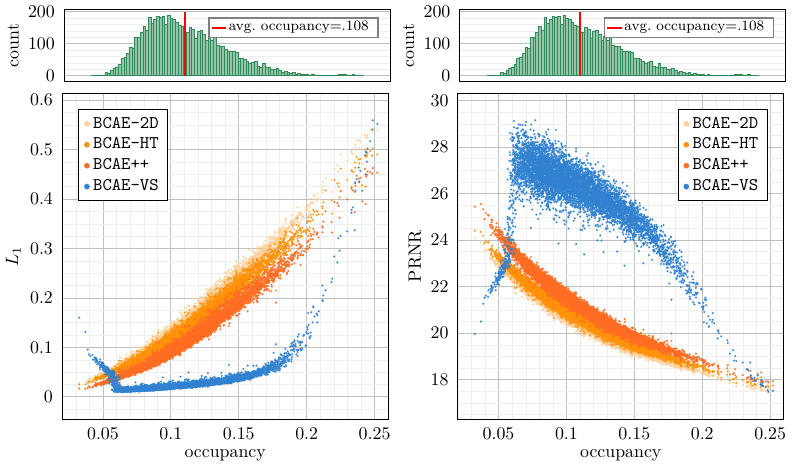}
    \fi 
    \caption{\textbf{Reconstruction $L_1$ error (left) and peak signal-to-noise ratio (PSNR, right) as a function of occupancy for dense {\bcae} models.}}
    \label{fig:metric_reg}
\end{figure}

\begin{figure}[ht]
    \centering
    \ifeditingfigure
        \tikzsetnextfilename{metric_clf}
        \def\scale{.86}
        \input{fig/metric_clf}
    \else
        \includegraphics{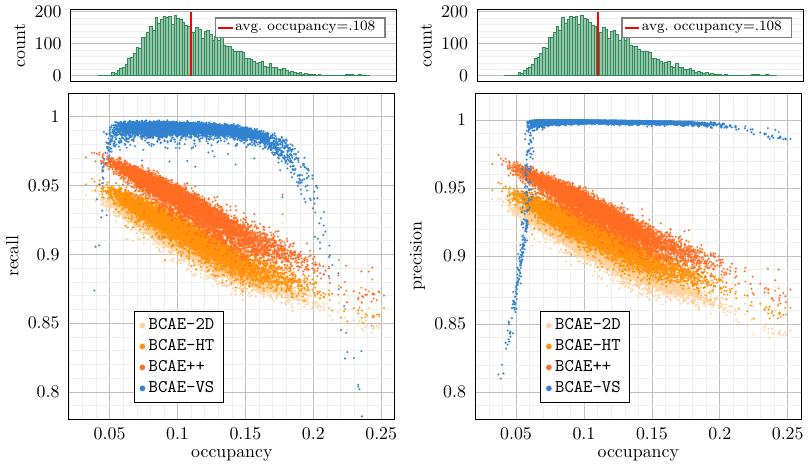}
    \fi 
    \caption{\textbf{Reconstruction recall (left) and precision (right) as a function of occupancy for dense {\bcae} models.}}
    \label{fig:metric_clf}
\end{figure}

\begin{figure}[ht]
    \centering
    \ifeditingfigure
        \tikzsetnextfilename{reconstruction}
        \resizebox{\textwidth}{!}{\input{fig/reconstruction}}
    \else
        \resizebox{\textwidth}{!}{\includegraphics{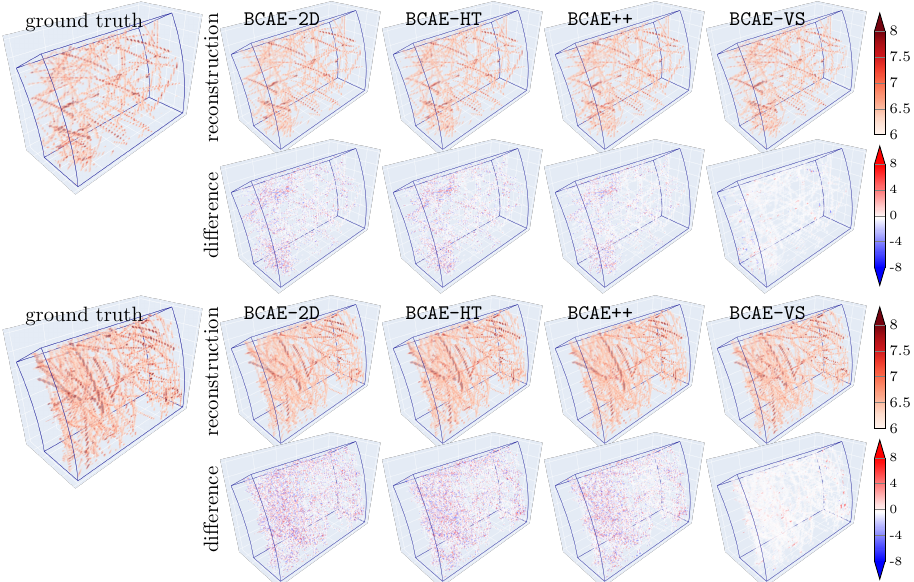}}
    \fi 
    \caption{\textbf{Visualization of reconstruction performance for dense {\bcae} models.} 
    }
    \label{fig:reconstruction}
\end{figure}

Since there is no value below $6$ in the input log ADC, we apply a zero-suppression at $6$ to the reconstruction produced by \modelname. The reason that we did not use the regression output transform approach (mapping all regression output to above $6$) introduced in dense {\bcae}s~\cite{huang2023fast} is because it makes the training of \modelname unstable.

We measured the reconstruction accuracy in $L_1$, $L_2$, peak signal-to-noise ratio (PSNR), recall, and precision and listed the result in \autoref{tab:main_result}. We computed reconstruction from compressed data saved in half-precision (float$16$). Here, recall is defined as the fraction of signals that get a positive value in the prediction, and precision is defined as the fraction of true signals over all voxels that has a positive prediction. To compare the compression ratio of the \modelname with those of the dense \bcae models, we use the convention detailed in \autoref{subsec:compression_ratio}. The encoder size is measured by trainable parameters. The throughput is measured by the number of TPC wedges processed per second.

We also plot the dependence of reconstruction accuracy on occupancy in \autoref{fig:metric_reg} for $L_1$ and PSNR and in \autoref{fig:metric_clf} for recall and precision. We can observe that the reconstruction accuracy metrics ($L_1$, recall, and precision) of the dense \bcae models depend linearly on occupancy with a strong correlation. On the contrary, in a wide range of occupancy where the TPC wedges are the most populated, the sparse model \modelname's performance is correlated weakly with occupancy and maintains a superior performance. However, towards the two ends of the occupancy distribution, the \modelname starts to fail. From \autoref{fig:metric_clf}, we can observe that the failure of \modelname is driven by a drastic decline in recall for extremely high-occupancy input and by precision for low-occupancy ones. 

In \autoref{fig:reconstruction}, we compare {\bcae}s' reconstruction performance qualitatively on two TPC wedges. The difference is calculated by subtracting reconstruction from the ground truth. From the comparison, we can make two observations. First, dense {\bcae}s (\bcaetwod, \bcaeht, \bcaeplus) have difficulty reconstructing a track at its correct location while \modelname conserves the track location even in densely packed regions. This is evident with the cloud of blue and red dots and the lack thereof in the plot of difference. Second, dense {\bcae}s under-perform in input region with higher occupancy while \modelname maintains accurate reconstruction over a wider range of occupancy.

\subsection{Compression Ratio}
\label{subsec:compression_ratio}
The compression ratio of the algorithm depends on data precision and the format of storage of the input and the code. 

On the input side, raw ADC values from TPC are collected as $10$-bit integers. However, since most commonly used computing platforms cannot handle $10$-bit integers, we consider input ADC values as $16$-bits floats. For the input size, we follow the convention in the dense {\bcae}s works~\cite{huang2023fast, huangTPCCompression} and define it as the size of the zero-padded regularly shaped tensors.

With respect to the precision of the code, although we train all models in full precision (float$32$), it is shown in~\cite{huang2023fast} that the difference in accuracy between reconstruction from full-precision compressed data and those down-cast to float$16$ is negligible. Similar observation can also be made for \modelname according to \autoref{tab:compare_full_half}. Hence, compressed data is saved in float$16$ format for all models.

\begin{table}[ht]
    \centering
    \caption{Reconstruction accuracy with full (float$32$) and half (float$16$) precision code.}
    \begin{footnotesize}
        \begin{tabular}{cccccc}
            \toprule
            precision & $L_1$ & $L_2$ & PSNR & recall & precision \\
            \midrule
            half & $.028394$ & $.088652$ & $26.000$ & $.988447$ & $.995573$ \\
            full & $.028415$ & $.088818$ & $25.996$ & $.988419$ & $.995569$ \\
            \bottomrule
        \end{tabular}        
    \end{footnotesize}

    \label{tab:compare_full_half}
\end{table}

Since a dense encoder output a code as a regularly shaped tensor, the compression ratio of a dense \bcae can be calculated directly as the ratio of the input size to the code size. However, since the code produced by the sparse encoder of \modelname is a sparse tensor, to compute a compression ratio comparable to those of the dense models, we need the following conversion. Assume that the Coordinate List (\texttt{COO}) format is used for saving the output from the sparse encoder of \modelname. We need to save the location and value of all entries in $v$ whose corresponding value in $p$ is larger than a (random) threshold. Since a TPC wedge has shape $(16, 192, 249)$, we need $4$ bit to specify the index of in the first dimension, and $8$ bits each for the second and last dimensions. Hence we need $(4 + 8 + 8) + 16=36$ bits to save one entry in $v$. Suppose the occupancy of the wedge is $o$ and a fraction $k$ of the entries in $v$ will be retained, the compression ratio $C$ for this wedge can be calculated using the following formula 
\begin{equation}
    C = \frac{16\text{bits} \times \text{input size}}{36\text{bits} \times \text{input size} \times {o} \times{k}} = \frac{4}{9ok}
    \label{eq:compression_ratio}
\end{equation}

\begin{figure}
    \centering
    \ifeditingfigure
        \tikzsetnextfilename{prob_cluster}
        \resizebox{\textwidth}{!}{\input{fig/prob_cluster}}
    \else
        \resizebox{\textwidth}{!}{\includegraphics{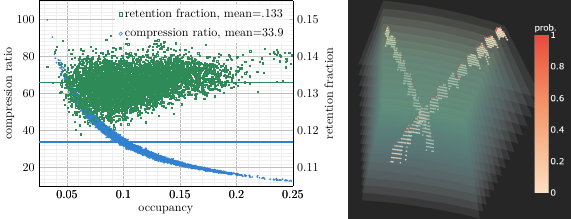}}
    \fi 
    \caption{\textbf{Compression ratio, retention fraction (left), and importance assignment to the signal voxels (right).}}
    \label{fig:compr}
\end{figure}

Over the test split of the dataset, on average a fraction of $.133$ of the signal voxels is retained for each wedge. The averaged compression ratio is $33.9$ with wedge-wise compression ratio calculated using \autoref{eq:compression_ratio}. The left panel of \autoref{fig:compr} shows compression ratio and retention ratio as a function of occupancy. While the dependency of retention on occupancy is relatively weak, \modelname tends to retain more signal voxels at higher occupancy. This behavior is desirable, as more voxels need to be preserved for accurate reconstruction when trajectories are closer together. The compression ratio, on the other hand, exhibits a clear inverse relationship to occupancy. These observations demonstrated \modelname's ability to dynamically adjust key-point identification and achieve a variable compression ratio. 

We also visualize the importance score assignment by the encoder of \modelname on isolated tracks in the right panel of \autoref{fig:compr}. We can see that the assignment is polarized -- most of the probabilities are close to zero and only a small fraction of signal voxels along the boundary of a track are assigned with a high importance score.

\subsection{Throughput}
\label{subsec:throughput}

\begin{figure}[ht]
    \centering
    \tikzsetnextfilename{throughput}
    \includegraphics[scale=1]{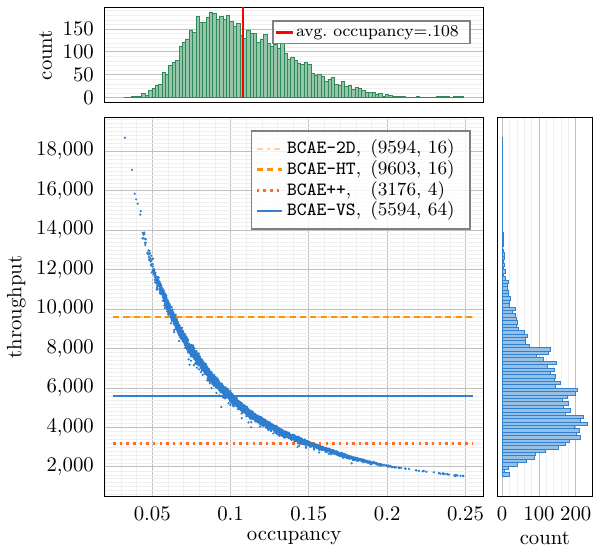}
    \caption{\textbf{Throughput comparison.} The lines of \bcaeht and \bcaetwod overlap. The numbers in the parenthesis are the TPC wedges compressed per second and the batch size used to obtain the throughput.
    } 
    \label{fig:throughput}
\end{figure}

Unlike dense convolutions, the inference speed of sparse convolution depends on occupancy. As discussed in \autoref{subsubsec:sparse_bcae_model}, lower occupancy leads to fewer matrix multiplications. Therefore, in an occupancy regime where the speed gain reduced matrix multiplications outweigh the overhead of computing the kernel map, \modelname may achieve lower compression latency. To test this hypothesis, we conducted a throughput study using a \chip graphic card with \driver and \torch compiled with \cuda\footnote{A fix build of \mink for \cuda can be found at \url{https://github.com/NVIDIA/MinkowskiEngine/pull/567}} and demonstrate the result in \autoref{fig:throughput}. Here, throughput is measured as the number of TPC wedges compressed per second. 

The throughput for each dense {\bcae} is averaged from $1000$ batches with a warm-up of $10$ batches. We evaluated throughput under the assumption that the input data is already stored in GPU memory. \autoref{fig:throughput} presents the peak throughput of each dense model, with the corresponding batch size that achieved the peak noted in the legend. All the dense {\bcae}s are compiled using \texttt{torch.compile} to optimize the inference speed. 

For \modelname, each test sample was repeated 64 times to generate the input. 
We assume the input batch of TPC wedges is already in GPU memory in a sparse-encoded format, ready for direct processing by the sparse network. This assumption is reasonable as the TPC data is collected and transferred in sparse encoding through the data acquisition pipeline. 

The throughput is calculated on the test split of the dataset consisting of $6288$ TPC wedges. 
Although \modelname is slower than \bcaetwod and \bcaeht on average, it is significantly faster in the lower occupancy regime. Moreover, sparse models can process sparsely encoded TPC data directly, eliminating the overhead of padding sparse encoding to a full tensor necessitated by a dense model. 
In many operational modes of collider tracking detectors, the occupancy can be significantly more sparse, e.g.~on the order of $10^{-3}$-$10^{-2}$ for the TPC in the sPHENIX experiment during proton-proton collisions, which is much lower than in the high-pileup central Au+Au dataset studied in this work. The advantages of sparse convolution become increasingly pronounced in such conditions.
\section{Conclusion}
We introduced \modelname, an algorithm designed to enhance the compression of sparse 3D TPC data from high-energy particle accelerators. \modelname offers a variable compression ratio based on data occupancy, improving both compression efficiency and reconstruction accuracy. Our results show a $75\%$ increase in reconstruction accuracy with a $10\%$ higher average compression ratio compared to the leading \bcae model. By utilizing sparse convolution, \modelname achieves an outstanding throughput within the operational occupancy range of sPHENIX. While the current model demonstrates promising performance, several limitations have been identified. We offer recommendations for future work to address these constraints and further enhance the model's capabilities.

\begin{itemize}
    \item \textit{Exact control of retention fraction:} Although hyperparameters such as $\lcomp$ and $\lowerprob$ (see~\autoref{subsubsec:losses}) can influence \modelname's the retention fraction, we lack a direct control of it. A potential solution could involve thresholding by quantile of the importance score output $p$ (see~\autoref{subsubsec:sparse_bcae_model}), allowing a fixed fraction of signals to be saved.
    \item \textit{Performance at extreme occupancies:} While \modelname performs well for typical occupancy levels in the dataset, it experiences a significant decrease in recall at high occupancy and in precision at low occupancy. For deployment, we need an algorithm that remains reliable even during extreme events.
    \item \textit{Precision and quantization for throughput:} The sparse convolution library \mink does not support half-precision (float16) operations. As demonstrated in~\cite{huang2023fast}, certain hardware can achieve substantial speedups with negligible performance loss using half-precision computation. Future work will explore more aggressive quantization schemes, such as int8, to further speed up inference. 
    \item \textit{Hardware support for sparse convolution:} Unlike dense convolution, sparse convolution engines are not commonly supported by innovative AI hardware accelerators. This poses additional challenges for deploying \modelname on such hardware compared to dense convolutional models.
    \item \textit{Downstream task:} Currently, the performance of \modelname is evaluated with standard classification/regression metrics. However, being a compression algorithm of ADC readout, \modelname's performance needs to be tested on downstream tasks such as trajectory reconstruction and identification, which will be part of our future work.
    \item \textit{Noisy TPC data from real experiment:} In this study, we used simulated collision data. However, the TPC data from real experiments will contain noises caused by, for example, common ground noise pickup. Hence, \modelname needs to be retrained and tested with realistic noisy conditions in the actual TPC data and adjusted to incorporate noise-filtering mechanisms.  
\end{itemize}

\section*{Acknowledgment}
We thank the sPHENIX collaboration for access to the simulated dataset, which was used in the training and validation of our algorithm. We also thank the collaboration for the valuable interaction with the sPHENIX collaboration on this work. This work was supported by the Laboratory Directed Research and Development Program of Brookhaven National Laboratory, which is operated and managed for the U.S. Department of Energy Office of Science by Brookhaven Science Associates under contract No. DE-SC0012704.

\appendix
\section{\modelname neural network}
\label{app:network}

The \modelname network consists of a sparse encoder implemented with sparse convolution provided by the NVIDIA \mink library
and two decoders implemented with normal dense convolution. The input to the neural network is one TPC wedge treated as a $4$D array with $1$ channel and $3$ spatial dimensions.  

The sparse encoder has five 3D convolution layers, all with $2$ output channels, kernel size $3$, stride $1$, and no padding. The dilation parameters of the convolutions are $1, 2, 4, 2, 1$. There is a rectified linear unit (\texttt{ReLU}) activation between two successive convolutional layers. The output activation is \texttt{Sigmoid}. 

The two decoders share the same structure. Each one of them has eight (dense) convolutional blocks with one convolution (kernel size $3$, padding $1$, stride $1$, dilation $1$) followed by a leaky \texttt{ReLU} with negative slope $.1$. The number of output channels are $(16, 16, 16, 8, 8, 8, 4, 2)$. A final $1\times1$ convolution maps the number of channels back to $1$. The regression decoder $\dreg$ has no output activation while the segmentation decoder $\dseg$ has \texttt{Sigmoid} as output activation. 

No normalization is used in the \modelname model. 

\section{\modelname training}
\label{app:training}

The training is done for $100$ epochs with a batch size of $4$ and $2000$ batches per epoch. For optimization, we used the Adam optimizer with decoupled weight decay (\texttt{AdamW}) with an initial learning rate $0.001$, $\paren{\beta_1, \beta_2} = (0.9, 0.999)$, and weight decay $0.01$. We kept the learning rate as initialized for the first $20$ epochs and decreased it to the $95\%$ of the previous value every $20$ epochs thereafter.

\bibliographystyle{elsarticle-num} 
\bibliography{bib/bib}

\end{document}